\documentclass[
 reprint,
 amsmath,amssymb,
 aps,
]{revtex4-2}

\usepackage{graphicx}
\usepackage{dcolumn}
\usepackage{bm}
\usepackage{xcolor}

\usepackage{hyperref}

\begin{document}

\title{Cyclic perturbations encode memories in trained neural networks without loss of performance}

\author{Max Kerr Winter$^{1,2}$}
\author{Liesbeth M. C. Janssen$^{3,4}$}
\affiliation{
$^1$Amsterdam Institute for Life and Environment, Faculty of Science, Section Systems Ecology, Vrije Universiteit Amsterdam, Amsterdam 1081 HV, The Netherlands\\
$^2$Infomatter Department, AMOLF, Amsterdam 1098 XG, The Netherlands\\
$^3$Soft Matter and Biological Physics, Department of Applied Physics, Eindhoven University of Technology, P.O. Box 513, 5600 MB Eindhoven, The Netherlands \\
$^4$Institute for Complex Molecular Systems, Eindhoven University of Technology,
P.O. Box 513, 5600 MB Eindhoven, The Netherlands \\
$^5$Institute for Theoretical Physics, Utrecht University, Princetonplein 5, 3584 CC Utrecht,
The Netherlands
}
\date{\today}

\begin{abstract}
Many disordered materials exhibit a remarkable form of memory: they retain information about the history of cyclic perturbations. Whether this phenomenon also occurs in artificial neural networks, whose mathematical structure and dynamics share many similarities with  disordered physical systems, is unknown. Here we show that cyclically perturbing a trained, overparameterized deep neural network encodes information in the structure of the weights that can subsequently be retrieved, thus retaining memory of the perturbation history. During each write cycle, all weights are displaced, after which the network relaxes through gradient descent. Repeated cycles organise the neural network into a state that is reversible for perturbations up to the write amplitude, allowing the encoded value to be read from the onset of irreversible weight motion. Multiple memories can also be encoded, although under deterministic writing only the largest persists at long times. Inspired by cyclically sheared suspensions, we introduce a novel noisy-write protocol that preserves all encoded memories at long times. Remarkably, neither deterministic nor noisy memory writing measurably affects the network's performance on unseen data from its original learning task. Our results establish cyclic memory as a phenomenon shared by disordered physical systems and deep neural networks, and show that networks can exploit as yet unexplored memory capabilities without compromising their original function.
\end{abstract}

\maketitle

\section{Significance Statement}
Building on the immense success of machine learning in recent years, the field of physical learning has made great progress imbuing materials with information storage and processing properties. For example, repeated deformation of a disordered solid organizes its microscopic components so that a previous deformation amplitude can be recovered. We show the same form of memory in trained deep neural networks. Repeatedly perturbing and relaxing a network’s weights encodes one or more amplitudes in a readable configuration. As such, our work uncovers a novel capability of deep neural networks that does not sacrifice existing functionality. This is a rare example of computational phenomena in physical materials inspiring a discovery in artificial neural networks, as opposed to the other way around.

\section{Introduction}
Memory effects -- where the state or dynamics of a system depends on its history -- play a significant role in a vast number of systems across the natural sciences and engineering. In recent years, an intriguing set of phenomena have emerged whereby memories (i.e.\ sequences of values) can be written and read from systems of disordered particles, without imposing any obvious order on them \cite{Keim2011,Keim2013,Keim2019,Paulsen2025}. This result is particularly striking given that established methods for recording information, e.g.\ written text, or bits encoded on a hard drive, are very deliberately not disordered. \\
 
How does reading and writing memories in disordered systems work in practice? Seminal work by Keim and Nagel \cite{Keim2011} demonstrated that a random suspension of particles can encode memory by subjecting it to repeated cyclic shear; the shear amplitude $A$ serves as the write amplitude, or memory. 
Initially, a shear cycle will result in many particle collisions, causing the particles to rearrange and hence fail to return to their original positions once the shear cycle is complete. However, after many shear cycles, the particles will be organised into positions where they no longer collide. Once all particles avoid collisions, the system exists in a steady state when viewed stroboscopically (i.e.\ once per cycle). This state encodes the write amplitude $A$. The read protocol consists of single shear cycles at varying amplitudes $\tilde{A}$, applied to copies of the system on which the memory has been written. When $\tilde{A}\leq A$, all particles return to their original positions at the end of the shear cycle. When $\tilde{A}>A$, collisions will start to occur hence some particles will be displaced. Consequently, $A$ can be deduced from the inflection point on a plot of the number of displaced particles against $\tilde{A}$. Many variations on this basic set up have been studied, including reading/writing multiple transient memories \cite{Keim2011,Keim2013}, memory formation in dense suspensions \cite{Fiocco2014,Fiocco2015}, memory formation in adaptive networks \cite{Bhattacharyya2022}, Hopfield-like associative memory \cite{Guo2026}, and the effect of random driving \cite{Mungan2025}.

Deep neural networks (DNNs) appear far removed from particulate materials, yet the two systems
share many common aspects in their mathematical structure. 
In DNNs, the loss function forms a high-dimensional landscape, with the network weights the degrees of freedom. Training corresponds to the network following a trajectory in weight space that minimises, typically stochastically, the value of the loss. In the network-particles analogy,
weights play the role of particle coordinates and the loss function plays the role of the energy
\cite{Franz2016,Geiger2019,Baity-Jesi2019,Bahri2020,Ghio2024,Winter2025}. The ratio $P/N$ of
the number of training examples, $P$, to network parameters, $N$, has a role resembling particle density. At low $P/N$,
overparameterized networks can fit every training example and access extended regions of the loss landscape with
training loss $\mathcal{L}=0$. This is much like the way in which a dilute gas with finite range interactions has an energy approximately equal to zero. Increasing $P/N$ produces an underparameterized regime with nonzero loss,
analogous to the onset of interactions in a denser particle system \cite{Geiger2019}.

However, the analogy between DNNs and disordered materials has important limits. Overparameterized networks have no established counterpart of
particle caging, their zero-loss regions can extend over long distances, and stochastic gradient
descent and its related training algorithms produce nonthermal noise that is anisotropic, has a non-flat power spectrum, and varies in intensity over time
\cite{Annesi2023,Yang2022a,Kuhn2023,Smith2020,Simsekli2019,Soudry2016,Keskar2017,Wu2022,
Wu2018,Winter2025}. Furthermore, trained networks can generalize to unseen data -- a phenomenon 
with no obvious physical counterpart. These differences make it unclear whether memory formation in disordered materials can be translated
to a neural network, or whether the motion within a low-loss region required by disordered memory read/write protocols would preserve the computation
learned by the network.

Here we demonstrate that trained DNNs can acquire readable memories under cyclic weight
perturbations. A deterministic protocol stores one long-lived amplitude and a second transient
amplitude. A modified noisy protocol stabilizes both amplitudes over the observation time.
Throughout, the networks retain their performance on unseen data. The memories therefore occupy
degrees of freedom that are available within a low-loss region while leaving the network's
original predictive function intact. As such, this work constitutes an unforeseen generalisation of the disordered memory concept to a new class of systems, as well as identifying a novel and as yet unexploited capability of deep neural networks.

\section{Results}

\subsection{Cyclic driving produces a readable memory}

\begin{figure}[h]
\includegraphics[width=8.7cm]{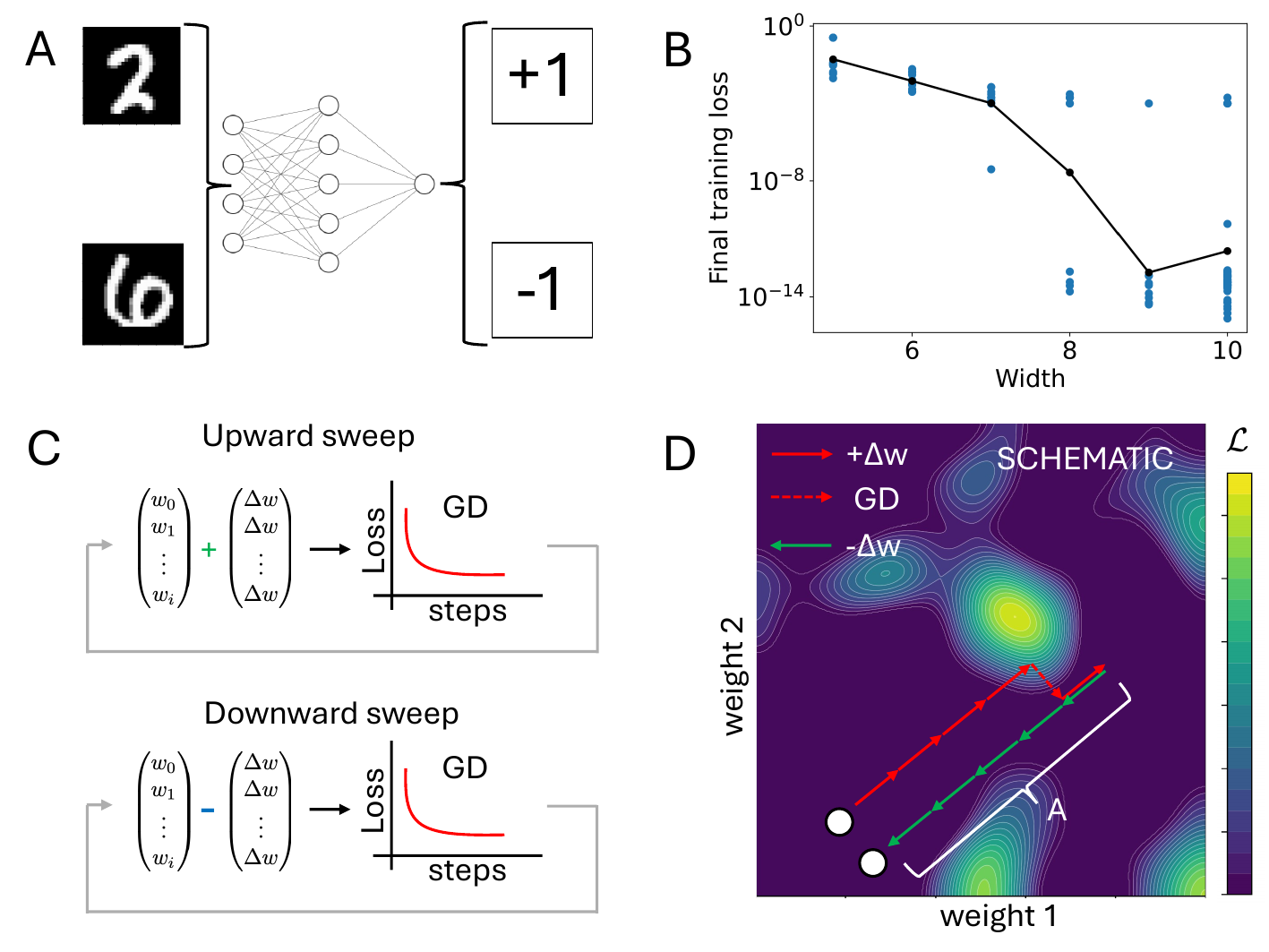}
\caption{
(A) Networks classify MNIST digits 2 and 6, assign labels $+1$ and $-1$, respectively.
(B) Final training loss as a function of hidden-layer width. Networks begin to reach
$\mathcal{L}\approx0$ near width 8, identifying the underparameterized-to-overparameterized
transition. Some wider networks remain at high loss because of their initial conditions and
stochastic training trajectories. (C) The write protocol consists of upward and downward weight
sweeps, with gradient-descent relaxation after each step. (D) Schematic trajectory through the
loss landscape. The initial and final states are indicated by white circles; upward and downward sweeps are depicted by red
and green arrows, respectively. Relaxation on a non-flat region (dashed arrow) prevents return to the
initial state.
}
\label{fig:fig1}
\end{figure}

As a proof of principle, we use the benchmark test of digit recognition from the standard MNIST dataset \cite{MNIST_dataset}, and ask whether a network trained on this task could exhibit memory effects analogous to those seen in particulate systems. For simplicity, we restrict ourselves to a binary supervised classification task where fully connected networks are trained to distinguish the MNIST digits 2 and 6. The images are reduced to 10 principal components, following the approach of \cite{Geiger2019} and \cite{Winter2025}, and networks with three
hidden layers are trained using the quadratic hinge loss. Figure \ref{fig:fig1}B shows that as hidden-layer width increases, the
final training loss drops by several orders of magnitude, marking entry
into the overparameterized regime. Reaching the overparameterised state is however dependent on a network's initial conditions and stochastic training trajectory. Hence, out of 50 width-10 networks (410 parameters), 41 achieve the low loss values characteristic of overparameterisation, defined here as $\mathcal{L}<10^{-8}$. In the remainder of this work, we exclusively use these 41 width-10 networks.

Our goal is to imprint a memory value, $A$, onto neural networks that have already been trained on the above digit recognition task, in such a way that we can recover the memory later. To do so, we introduce a cyclic protocol for DNNs inspired by the cyclic shearing of particle suspensions. Briefly, we displace every network weight in equal increments along a fixed
direction until the cumulative displacement reaches $A$. Gradient descent relaxes the network
after every increment. This is the upward sweep depicted in Fig.\ \ref{fig:fig1}C. We then reverse the displacement and increment the weights downwards, again interspersing the increments with periods of gradient descent, as depicted in the downward sweep of the same figure. The process of writing a memory consists of repeated applications of an upward sweep followed by a downward sweep. If the trajectory in weight space of an upward and downward sweep remains entirely on a flat
$\mathcal{L}=0$ region of the loss landscape, the network returns to its starting point. However, if the trajectory encounters a
nonzero-loss region, gradient descent displaces it, and so the completed cycle no longer
returns every weight to its initial value. Such an event is depicted schematically in Fig.~\ref{fig:fig1}D, and is analogous to a collision between particles during a cycle of shear in the case of memory writing in physical systems. Repeated write cycles thus move the network around in the loss landscape until a state is reached that is fully reversible up to the memory value, $A$.

\begin{figure}[h]
\includegraphics[width=8.7cm]{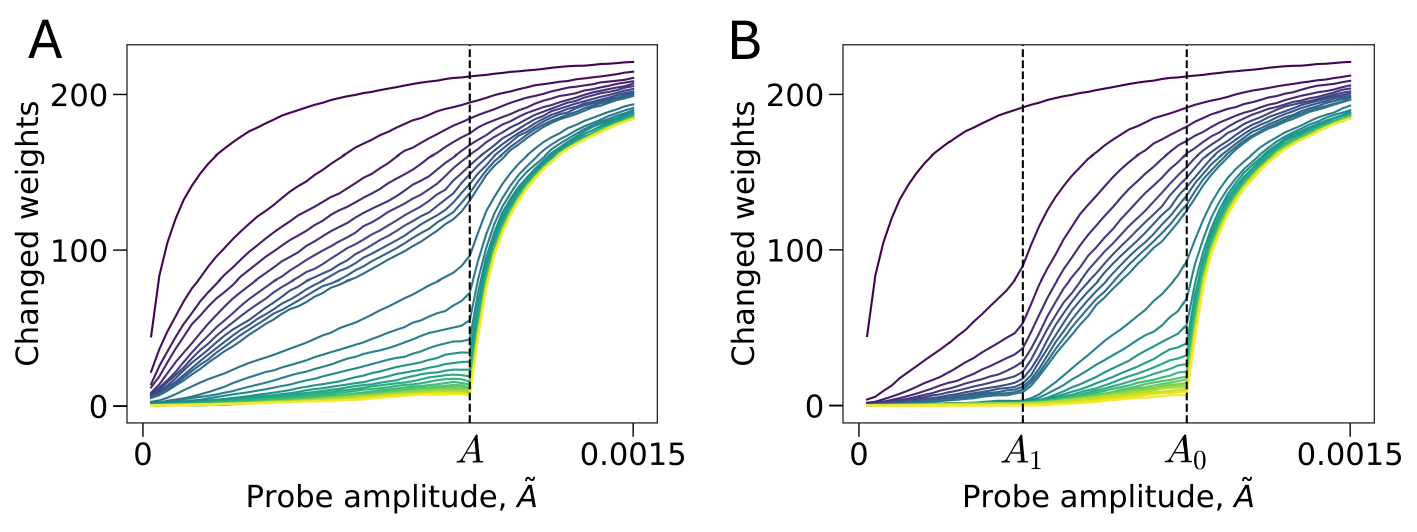}
\caption{
Memories written by deterministic cyclic perturbations. (A) Number of weights displaced by more
than $\epsilon=5\times10^{-6}$ after a read cycle of amplitude $\tilde{A}$. A memory written at
$A=0.001$ appears at the onset of weight motion. (B) Memories of $A_0=0.001$ and $A_1=0.0005$
appear as changes in slope. The smaller memory is erased after approximately 140 write cycles.
In both panels, color progresses from 0 write cycles (purple) to 200 cycles (yellow).
}
\label{fig:fig2}
\end{figure}

Subsequently, we \textit{read} the memory by applying one cycle of an upward and downward sweep to a copy of the written network at a
probe amplitude $\tilde{A}$, and then probing across many values of $\tilde{A}$. The readout is the number of weights displaced by more than a
sensitivity $\epsilon$ after the read cycle. What does this look like in practice? Figure \ref{fig:fig2}A shows the number of displaced weights as a function of $\tilde{A}$, averaged over 41 independently trained networks, each of which has been subjected to the write protocol with memory value $A=0.001$. The individual curves correspond to applying the read protocol after increasing numbers of write cycles. As the number of write cycles increases, a sharp inflection point emerges at $\tilde{A}=A$, demonstrating that the memory value has indeed been stored in the structure of the DNN, and can be successfully read by our protocol. The inflection point demarcates the boundary between reversible and irreversible weight trajectories under the read protocol.

The same read/write protocols can encode more than one value. To demonstrate this, we alternate a cycle at $A_0=0.001$ with six
cycles at $A_1=0.0005$. Once again applying the read protocol and plotting the result in Fig.\ \ref{fig:fig2}B, both memory values appear as changes in the slope of the read curves. As expected, the smaller amplitude is transient, and disappears after
approximately 140 write cycles. This hierarchy mirrors the behaviour of multiple transient memories in cyclically
driven suspensions: repeated driving ultimately organizes the system around the largest
amplitude, erasing signatures of smaller amplitudes \cite{Keim2011,Keim2013}. As such, the results in Fig.\ \ref{fig:fig2} exactly reproduce the phenomenology of particulate systems, and constitute a clear demonstration of the existence of cyclic memory in DNNs.

\subsection{A novel noisy write protocol stabilizes multiple memories}

Thus far, we have considered memory writing protocols in DNNs for the case of fully deterministic dynamics. How does the presence of noise change this picture? A striking aspect of cyclic memory in particulate systems is that noise can, counterintuitively, \textit{stabilise} multiple memories, preventing smaller memories from being erased by the largest one. The reason for this is that the noise continually returns
particles to regions of space from which they were previously swept out \cite{Keim2011}. Collisions correspond to the system moving out of a zero energy state, hence in the energy landscape picture, noise repeatedly drives the system onto the hills surrounding zero energy basins. 

It is not immediately obvious if noise will also preserve multiple memories in DNNs. Indeed, we find that the most literal translation of this mechanism to DNNs, i.e.\ directly
adding Gaussian noise to the network weights, does not reproduce the stabilising effect seen in particulate systems. In fact, it produces the more naively intuitive result that memories are simply erased by the noise (Supporting
Information). This suggests that noise does not \textit{generically} promote memory retention in DNNs, but rather requires a specific mechanism through which it may help stabilise existing memories. 

To establish how noise can enhance memory retention in DNNs, we modify the loss during the write cycles so that previously flat regions are assigned
a weak, randomly varying gradient pushing the network toward the finite-loss boundaries (see Eq.\ \ref{eq:spring_loss} for the full definition). We call this modification
``spring noise'', due to its qualitative similarity to placing springs between randomly selected pairs of particles. The read protocol remains unchanged and deterministic.

\begin{figure}[h]
\includegraphics[width=8.7cm]{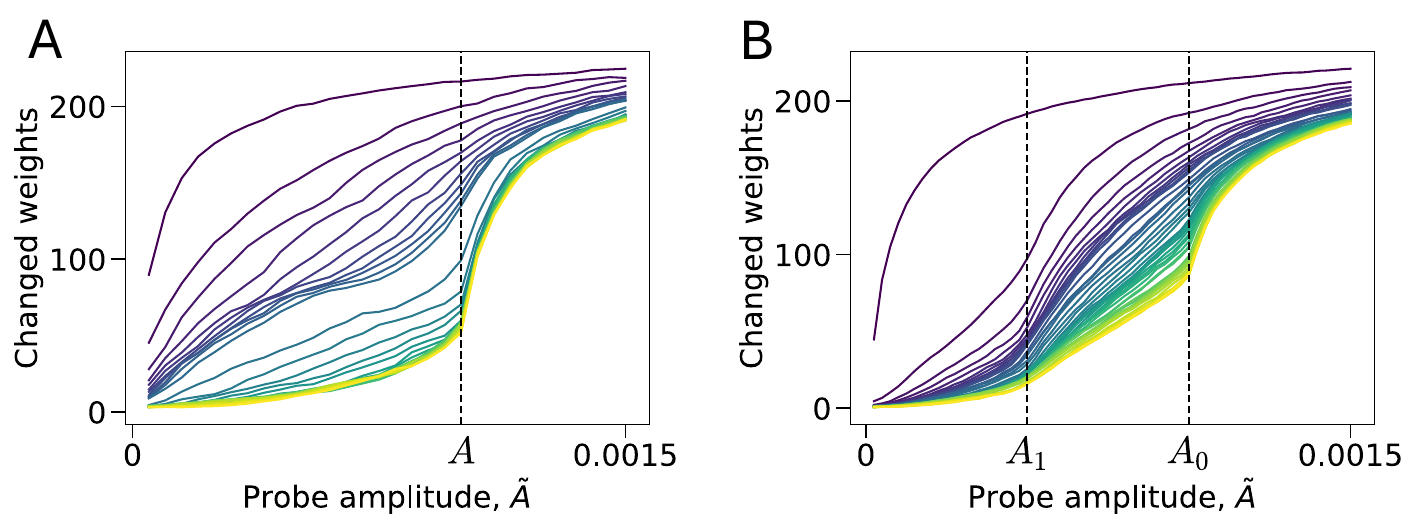}
\caption{
Memories written with spring noise. (A) A memory at $A=0.001$ remains visible, while weights
continue to move for $\tilde{A}<A$. Color progresses from 0 write cycles (purple) to 200 cycles
(yellow). (B) Memories at $A_0=0.001$ and $A_1=0.0005$ both remain visible after 360 cycles.
Color progresses from 0 cycles (purple) to 360 cycles (yellow). Both panels use
$\sigma=10^{-7}$ and $\epsilon=5\times10^{-6}$. See Eq.\ \ref{eq:spring_loss} for the definition of $\sigma$.
}
\label{fig:fig3}
\end{figure}

The effect of introducing spring noise during the write cycles of a memory written at $A=0.001$ is shown in Fig.\ \ref{fig:fig3}A. As with the deterministic case in Fig.\ \ref{fig:fig2}A, there is a large inflection at $\tilde{A}=A$, demonstrating that the memory has been successfully written and read. In contrast to Fig.\ \ref{fig:fig2}A, some weights continued to move for
$\tilde{A}<A$ after many write cycles, long past the point where the deterministic write protocol reaches a reversible state. This persistent subthreshold response is also seen in particles subjected to thermal noise, and is the network analog of continued particle collisions in previously swept out regions of space. 

We next apply the write protocol with spring noise to the case of multiple memories, with values
$A_0=0.001$ and $A_1=0.0005$, again applied in the same sequence as in the deterministic
experiment of Fig.\ \ref{fig:fig2}B. The spring noise result is shown in Fig.\ \ref{fig:fig3}B, where both memories remain visible as abrupt gradient changes in the readout curves after 360 write cycles. The smaller memory
therefore persists for substantially longer than its deterministic counterpart, which is
erased after approximately 140 cycles. Consequently, Fig.\ \ref{fig:fig3}B demonstrates the effectiveness of spring noise at recovering the phenomenology of noisy particle suspensions. The existence of a write/read protocol capable of storing multiple persistent memories in DNNs is crucial for any practical application of this phenomenon.

\subsection{Memory writing preserves predictive performance on unseen data}

\begin{figure}[h]
\includegraphics[width=8.7cm]{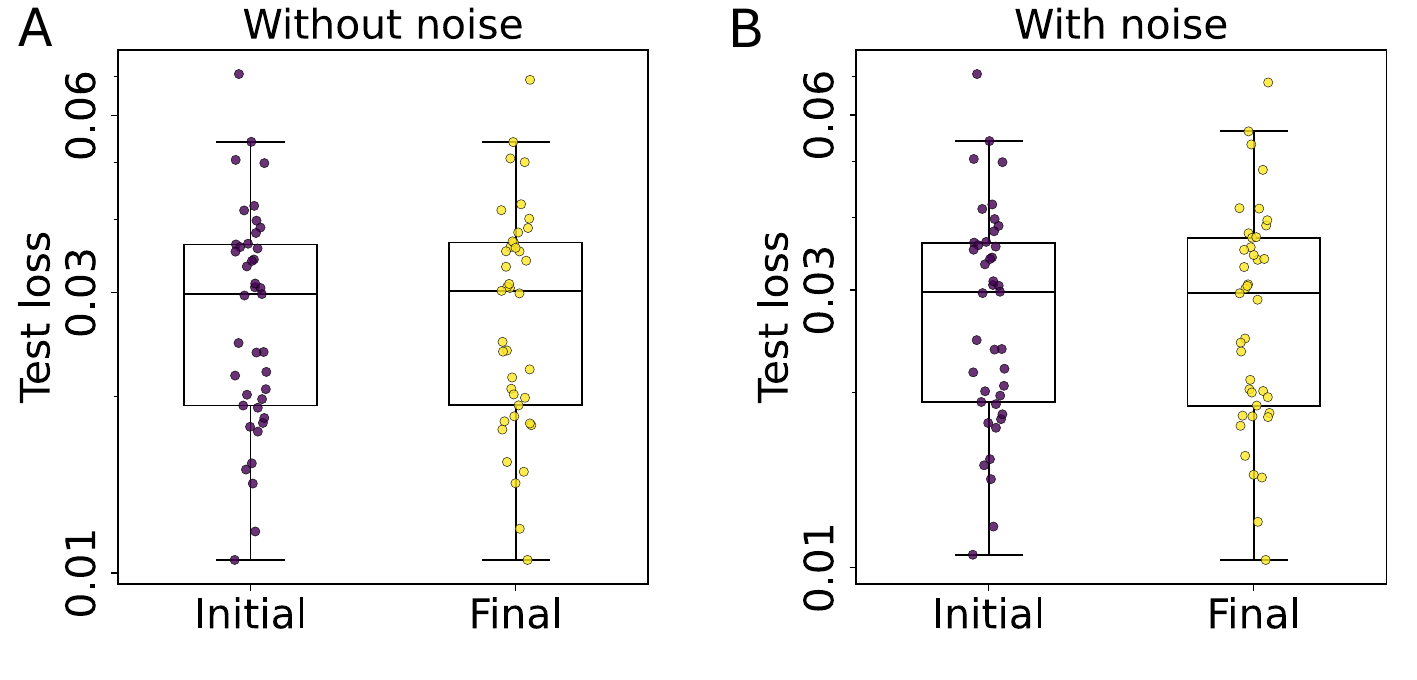}
\caption{
Effect of memory writing on predictive performance. Test-loss distributions are shown before
and after 200 cycles writing a memory at $A=0.001$. (A) Deterministic write protocol.
(B) Spring noise protocol with $\sigma=10^{-7}$. Neither protocol produces a discernible change
in test loss.
}
\label{fig:fig4}
\end{figure}

A major distinction between a neural network and a disordered material is that the network can perform a learned computation. This raises the important question of what effect, if any, writing and reading memories by cyclic perturbation has on the DNNs' performance on the task it was originally trained to perform. We therefore compare the networks' loss values on unseen test data from the original digit classification task before and after writing a memory. As shown in Fig.\ \ref{fig:fig4}, neither the deterministic protocol nor the spring noise
protocol produces a discernible change in the distribution of test losses. Thus, within the amplitudes and times studied, cyclic driving reorganized
the network weights enough to encode additional information without measurably degrading
generalization on the original classification task. In other words, the predictive performance on unseen data is not compromised at all, and yet the DNNs are enriched by additional memory capabilities.

\section{Discussion}

Our results establish a form of memory in trained DNNs that is distinct from the associative
memories of Hopfield networks and the persistent dynamical states of recurrent networks. A
Hopfield memory corresponds to an attractor, so storage capacity is related to the number of
stable minima \cite{Hopfield1982,Amit1985a}. Here an amplitude is represented instead by the
distance from the network's state to a finite-loss boundary along a prescribed perturbation
path. Capacity is therefore controlled by the geometry of a low-loss region rather than by the
number of minima.

This geometric distinction may help expose where the analogy between neural networks and
disordered materials succeeds and where it fails. In particulate systems, zero-energy regions
are constrained by nearby collisions and by boundary conditions. Such systems are typically isotropic, hence collisions occur when any particle moves in any direction by roughly one interparticle distance. By contrast, weights in an
overparameterized network can diffuse for long times without encountering a barrier
\cite{Annesi2023,Winter2025}. Furthermore, the weights of our DNNs have free boundary conditions, hence Gaussian noise is likely to result in an effect similar to a gas expanding, with fewer and fewer interactions as the particles move apart. 

These differences between DNNs and particles may explain why Gaussian weight noise does not
stabilize multiple memories: unconstrained diffusion can carry the network away from finite-loss
boundaries rather than repeatedly returning it to them. Spring noise is designed to restore
that return and successfully recovered the phenomenology of thermally driven particle systems, despite being dissimilar to thermal fluctuations. The importance of basin shape does, however, suggest a potentially important difference between memory formation by cyclic perturbations in DNNs compared with physical systems. If DNNs do indeed have much larger basins, as suggested above, then it could be that they will have a much higher memory capacity than a comparable set of particles.

The preservation of test performance shows that movement within a low-loss region can alter a
network's response to parameter perturbations without changing its measured predictive
function. This is an important condition for establishing cyclic memory as a useful addition to established DNN functionality. Furthermore, it moves DNNs a small step closer to the biological scenario of information storage and processing occurring in the same structures, as opposed to the standard computer architecture separating the two. 

Determining how many independent
amplitudes or perturbation directions can be stored, how capacity scales with network size, and
whether analogous memory phenomena exist in the underparameterized regime are immediate open questions.
More broadly, the results suggest that memory formation by cyclic perturbation could be a quite 
general phenomenon in disordered systems, not just restricted to particles, or the charge-density 
waves which the particle work was inspired by \cite{Coppersmith1997,Povinelli1999}. Perhaps the 
analogy could be extended even further afield, to disordered ecosystems \cite{Altieri2021} and beyond. 

\section{Materials and Methods}

\subsection{Data and network preparation}

We selected images of digits 2 and 6 from MNIST and assigned labels $y=+1$ and $y=-1$,
respectively \cite{MNIST_dataset}. Principal-component analysis reduced each $28\times28$ image
to 10 components. The training set contained 5,000 images and the test set contained 1,000
images, with equal numbers of each digit in both sets.

The models were fully connected, feed-forward networks without bias parameters. Each network had
10 input units, three hidden layers of equal width with ReLU activations, and one output unit.
We varied hidden-layer width from 5 to 10 to locate the overparameterization transition. Networks
were trained for 10,000 epochs by stochastic gradient descent with learning rate 0.01 and batch
size 32. Memory experiments used all width-10 networks whose final training loss was below
$10^{-8}$.

For $P$ training examples, the quadratic hinge loss, $\mathcal{L}$, was
\begin{align}
\mathcal{L}
&= \frac{1}{P}\sum_{i=1}^{P}\ell(f_i,y_i), \label{eq:quad_hinge_loss}\\
\ell(f_i,y_i)
&= \frac{1}{2}\max\left(0,\Delta(y_i,f_i)\right)^2,\\
\Delta(y_i,f_i)
&= 1-y_i f_i,
\end{align}
where $f_i$ is the network output for example $i$ \cite{Lee2013}. The flat zero of this loss provides extended
regions in which perturbations can be reversible.

\subsection{Deterministic write protocol}

Starting from trained parameters $\theta$, we added a scalar increment $\Delta w$ to every
weight, then performed $N_{\mathrm{GD}}=2,000$ full-batch gradient-descent steps at learning rate
$\eta=0.01$. We repeated the displacement and relaxation $A/\Delta w$ times, then reversed the
process by subtracting $\Delta w$ before each relaxation. The upward and downward sweeps formed
one write cycle. Unless stated otherwise, $\Delta w=2.5\times10^{-5}$. Manual checks on several
networks confirmed convergence within 2,000 relaxation steps.

Single-memory experiments repeatedly used cycles with $A=0.001$. For two-memory experiments, cycles
were applied in the sequence
\begin{align}
(A_0,A_1,A_1,A_1,A_1,A_1,A_1),
\end{align}
with $A_0=0.001$ and $A_1=0.0005$. Repeating the smaller amplitude was not required to create the
memory but accelerated the evolution of the low-amplitude part of the read curve.

\subsection{Read protocol}

Each read amplitude $\tilde{A}$ was tested on a separate copy of the written network. We applied
one upward and downward cycle using the same $\Delta w$, relaxation time, and learning rate as
in the write protocol. The memory readout was the number of weights whose final values differed
from their values before the read cycle by more than $\epsilon=5\times10^{-6}$. Changes in the
slope of this count as a function of $\tilde{A}$ identified written amplitudes.

\subsection{Spring-noise protocol}\label{sec:spring_noise}

During noisy writing, we replaced the per-example loss by
\begin{align}
\ell(f_i,y_i)
= \frac{1}{2}\max\left(\sqrt{m(f_i-y_i)},\Delta(y_i,f_i)\right)^2,\label{eq:spring_loss}
\end{align}
where $m\sim\mathcal{N}(0,\sigma)$ and the remaining definition of $\mathcal{L}$ was unchanged.
A new $m$ was sampled after each increment or decrement of $\Delta w$. This replacement gives
the nominally flat part of the loss a random gradient directed toward the quadratic region. We
used $\sigma=10^{-7}$ for the results in Fig.~\ref{fig:fig3}. All reads were performed without
noise. The value of $\sigma$ is somewhat arbitrary as there is no system defined energy scale. Instead, $\sigma$ must be chosen manually so as to be strong enough to preserve small memories, but not so strong as to eradicate all memories. Such a balancing act is similar to the process of choosing the sensitivity $\epsilon$ without a physical length scale.

\subsection{Gaussian weight-noise control}

\begin{figure}[h]
\includegraphics[width=8.7cm]{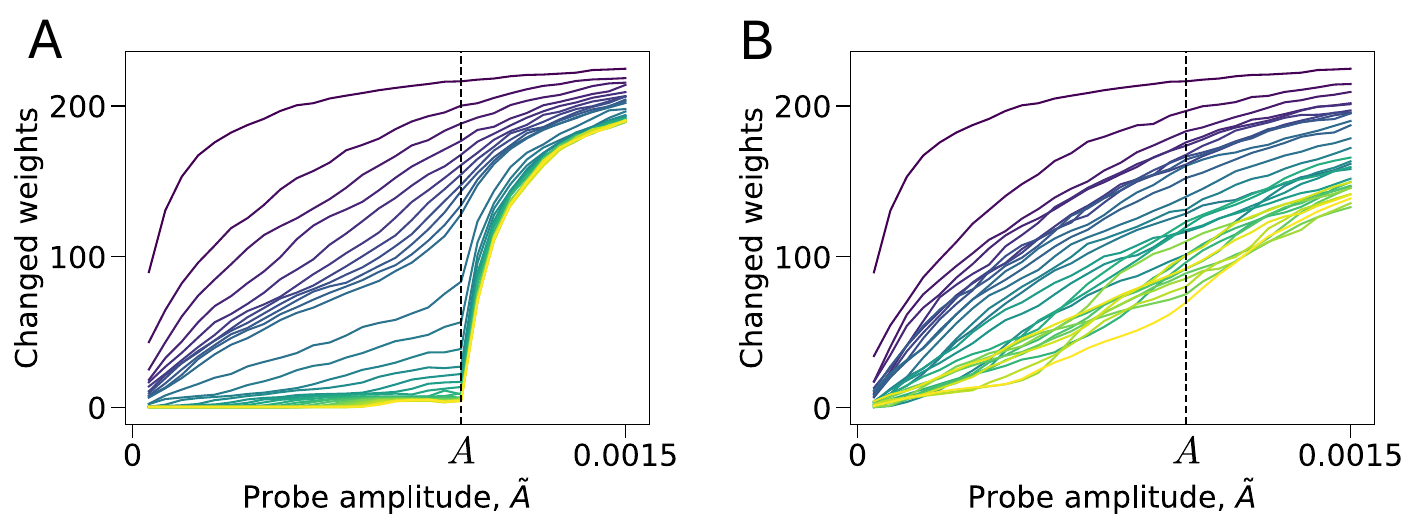}
\caption{
Gaussian weight-noise control. (A) At low noise levels, the number of changed weights remains zero below the writing
amplitude $A=0.001$. Parameters are $\epsilon=5\times10^{-6}$ and noise standard deviation
$10^{-6}$. (B) At high noise levels, the memory is simply erased. Parameters are $\epsilon=5\times10^{-6}$ and noise standard deviation
$10^{-4}$.
}
\label{fig:figS1}
\end{figure}

As a control, independent Gaussian random values were added directly to all weights after each
increment or decrement during writing. At standard deviation $\sigma=10^{-6}$, a memory written at
$A=0.001$ retained a flat response below $A$ (Fig.~\ref{fig:figS1}A). Smaller noise had no visible
effect, whereas larger noise removed the sharp response at $A$, as shown in Fig.\ \ref{fig:figS1}B, with $\sigma=10^{-4}$.

\section{Acknowledgments}
M.K.W. gratefully acknowledges funding from Stichting Ammodo. This work used the Dutch national e-infrastructure with the support of the SURF Cooperative using grant no. EINF-19321. 

\section{Author Contributions}
M.K.W. and L.M.C.J. designed the research; M.K.W. performed the research and analyzed the data; M.K.W. and L.M.C.J. wrote the paper.

\section{Competing Interest Statement}
The authors have no competing interests.

\section{Data, Materials, and Software Availability}
The code used to generate all of the data, results and plots in this paper is available at \url{https://github.com/mkerrwinter/cyclicMemory}.

\bibliographystyle{apsrev4-1}
\bibliography{cyclic_memory.bib}

\end{document}